\documentclass[reqno,12pt]{amsart}
\usepackage{amsmath, latexsym, amsfonts, amssymb}
\usepackage{mathrsfs,enumerate}
\usepackage{graphics,epsfig,color}

\usepackage{graphicx}
\usepackage{epstopdf}

\numberwithin{equation}{section}

\newcommand{\cG}{{\ensuremath{\mathcal G}} }

\newcommand{\cJ}{{\ensuremath{\mathcal J}} }

\newcommand{\cI}{{\ensuremath{\mathcal I}} }

\newcommand{\bbP}{{\ensuremath{\mathbb P}} }

\newcommand{\bbR}{{\ensuremath{\mathbb R}} }

\newfont{\indic}{bbmss12}

\title[Macroscopic fluctuation theory of local collisional dynamics]
      {Macroscopic fluctuation theory of local collisional dynamics}

\author[R.\ Lefevere]{Rapha\"el Lefevere}
\address{Laboratoire de Probabilit\'es
 et Mod\`eles Al\'eatoires (CNRS UMR 7599), Universit\'e Paris 7
 -- Denis Diderot, UFR Math\'ematiques, Case 7012, B\^atiment
 Chevaleret, 75205 Paris Cedex 13, France}
\email{lefevere\@@math.univ-paris-diderot.fr}

\begin{document}

\begin{abstract}
We explain why the macroscopic fluctuations of deterministic local collision dynamics should be characterized by a non strictly convex functional.
\end{abstract}



\maketitle
The study of thermal conduction properties of extended lattice Hamiltonian systems has received recently a lot of attention by theoretical and mathematical physicists alike.  The first issue is simply the derivation of the Fourier law, or equivalently the heat equation. On the other hand, an impressive amount of rigorous results concerning stochastic interacting particles system has been obtained over the last decades.  The diffusion equation has been derived in the hydrodynamic limit and the behaviour of the fluctuations of the macroscopic observables of the system under a space-time diffusive scaling has been derived rigorously .  In particular, the shape of the large deviations functional of the current and density of particles profile has been determined \cite{jona1,jona2,jona0,jona3,BD1,BD2,BD3,OM}.  This object may be seen as the out-of-equilibrium analogue of the equilibrium thermodynamic potentials.

The first example of local collision dynamics has been introduced in \cite{bunilive}.  In that model, particles are locked in cells having a particular shape ensuring strong chaotic properties for the dynamics of each particle within its own cell.  The cells are arranged so as to tile the plane and have a small opening so that particles in neighbouring cells may interact through elastic collisions. This was model taken up in \cite{GG} as a model for heat conduction when its boundaries are connected to heat baths.  The authors argue that in a weakly interacting regime and after a suitable time rescaling, the energy exchange between neighbouring cells is described by a stochastic dynamics analogous to the ones encountered in stochastic interacting particles.  This raised the hope of using the rigorous results obtained for those systems to derive Fourier law for deterministic dynamics. 
Unfortunately, the stochastic dynamics describing the exchange of energy does not satisfy the gradient condition and is therefore difficult to analyze. 
In this paper, we want to explain briefly how another approach to local collision dynamics allows to conjecture important features regarding the fluctuations (large deviations) of its macroscopic observables.  This was introduced and developed  in \cite{GL, LefevereZambotti1,LMZ1,LMZ2}.
\medskip
Consider $N$ particles  of unit mass with positions and momenta
$(\underline{\mathbf{q}}, \underline{\mathbf{p}}) \equiv
\big\{(\mathbf{q}_i, \mathbf{p}_i)\big\}_{1\leq i\leq N}$, with
$\mathbf{q}_i, \mathbf{p}_i \in \mathbb{R}^d$.  The positions are measured with respect to $N$ fixed centers located on a 1D lattice. The Hamiltonian $H$ takes
the form :
\begin{equation}
H(\underline{\mathbf{p}}, \underline{\mathbf{q}})
= \sum_{i=1}^N \left[\frac{\mathbf{p}_i^2}{2} +V(\mathbf{q}_i)+
U(\mathbf{q}_{i}-\mathbf{q}_{i+1}) \right],
\label{Hamilton}
\end{equation}
where the interaction potential $U$ is equal to zero inside a region $\Omega_U\subset\mathbb{R}^d$ with
smooth boundary $\Lambda$ of dimension $d-1$, and equal to infinity outside.
Likewise, the pinning potential $V$ is assumed to be zero inside a
bounded region $\Omega_V$ and infinity outside, implying that the motion of
a single particle remains confined for all times. The regions $\Omega_U$ and
$\Omega_V$ being specified, the dynamics is equivalent to a billiard in
high dimension.  A typical example of the dynamics we wish to consider is given by the figure below.  The circles move freely within their square cells and collide with each other when they both get sufficiently close to the hole located in the wall separating two adjacent cells.

\begin{figure}[thb]
\includegraphics[width = 1\textwidth]{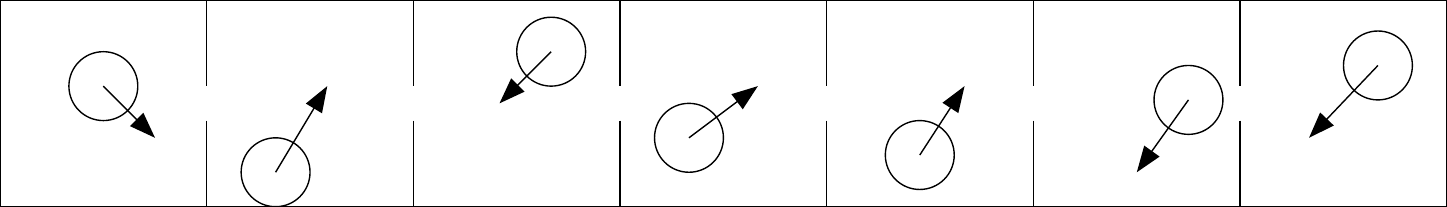} 
\caption{Simplified one dimensional aerogel dynamics}
\label{simple}
\end{figure}

Physically, those models describe aerogels, i.e. gels whose liquid components have been removed and replaced by atoms of gases.
One of the simplest example of this type of dynamics is the complete exchange model introduced in \cite{Prosen}.  Although it is expected to display anomalous thermal conduction properties, it has nice aspects that underline the general important features of local collision dynamics.
In the definition of the Hamiltonian (\ref{Hamilton}),  take $d=1$ and the potentials $V$ and $U$ given by
\begin{eqnarray}
\label{eq: well}
V(x)=\left\{
\begin{array}{l}
  +\infty\; {\rm if}\; |x|>b\\
  0\; {\rm if }\; |x|\leq b
\end{array}
\right. \qquad U(x)=\left\{
\begin{array}{l}
  +\infty\; {\rm if}\; |x|>a\\
  0\; {\rm if }\; |x|\leq a
\end{array}
\right.
\nonumber
\end{eqnarray}

Each particle on the lattice moves freely on a one-dimensional cell of size
$2 b$, changing directions at the boundaries. The interaction 
between a pair of particles acts when the difference between the positions
of the two particles reaches the value $a$, at which point they exchange
their velocities.

\noindent The motion of a given pair of particles at sites $i, i+1$ is described as the motion of a point particle on a two-dimensional billiard table described by $$\Omega=\left\{(x_1,x_2)\in {\bf R}^2,\, |x_1|\leq b,\, |x_2|\leq b,\,  |x_1-x_2|\leq a\right \}.$$ 

\begin{figure}[thb]
\includegraphics[width = .40\textwidth]{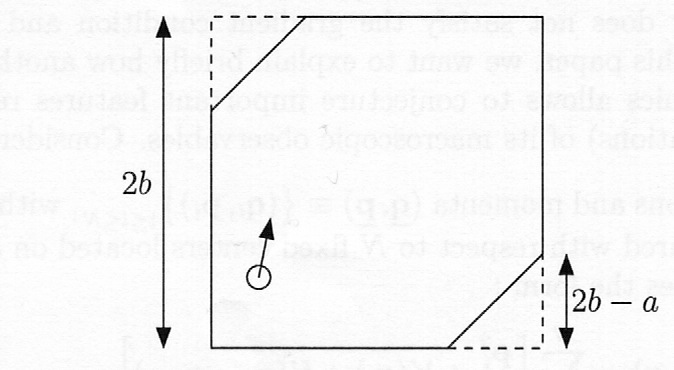} 
\caption{Billiard giving the motion of two particles in the complete exchange model}
\label{2dbilliard}
\end{figure}
It is straightforward to see that the energy of the particle $n$ at time $t$ is given by
$$
E(n,t)-E(n,0)=J(n-1,[0,t])-J(n,([0,t])
$$
where $J(n-1,[0,t])$ is the time-integrated current:
\begin{equation}
J(n,[0,t])=-\frac 1 2\sum_{k=1}^{C_n(t)}[p^2_{n+1}(\tau_n^k)-p^2_{n}(\tau_n^k)]
\label{current_hard}
\nonumber
\end{equation}
$(\tau_n^k)_k$ is the sequence of collision times between particles $n$ and $n+1$ and $C_n(t)$ is the number of collisions up to time $t$. For more general local collision dynamics, the current keeps this ``gradient"-like expression with a prefactor in front of the difference of kinetic energy, see \cite{GL} for a detailed expression.  This a fundamental difference with the expression of the current that is found in chains of oscillators interacting through ``soft" interactions (i.e interacting with a smooth potential $U$).  In those systems, the time integrated current is of the form 
\begin{equation}
J(n,[0,t])=-\frac 1 2\int_0^t(p_n(s)+p_{n+1}(s))U'(q_n(s)-q_{n+1}(s))ds,
\label{current_soft}
\nonumber
\end{equation}
when the particles move in a 1d potential wall. It is easy to see that in that case, the expectation of the instantaneous current (the integrand in the above expression) with respect to a local equilibrium measure is zero.  This is not the case with the expression in (\ref{current_hard}).

We explain now how to construct a stochastic dynamics that is simple to analyze and yet hopefully retains the essential features of the determinisitic models.
The basic idea behind the model is the following : when the system is large, each particle in a given cell sees its neighbouring particles as being embedded in a heat bath whose temperature changes slowly on a microscopic time scale, meaning that many collisions occur before the temperature of the bath is significantly affected.  Conceptually, one should note that it is a very different way (and a more generic one) of introducing randomness in the picture than the one advocated in \cite{GG}.  Here the randomness in the motion of one given particle appears as the result of the action of a large number of particles, not as an effect of the specific shape of the cell in which each particle moves.  Roughly speaking, we replace the walls through which the collisions between neighbours occur by stochastic thermal walls at fixed temperatures and let the particles travel back and forth between the thermal walls.  In fact, we dissociate two aspects of the dynamics : each particle carries some energy by travelling between the two sides of its cells and on the other hand, it acts on its neighbours as part of a huge thermal bath when colliding with them. In summary, we build a model made of tracers and scatterers, the latter modelling the actions of a heat bath at a given temperature. Pictorially, we get this :

\begin{figure}[thb]
\includegraphics[width = .90\textwidth]{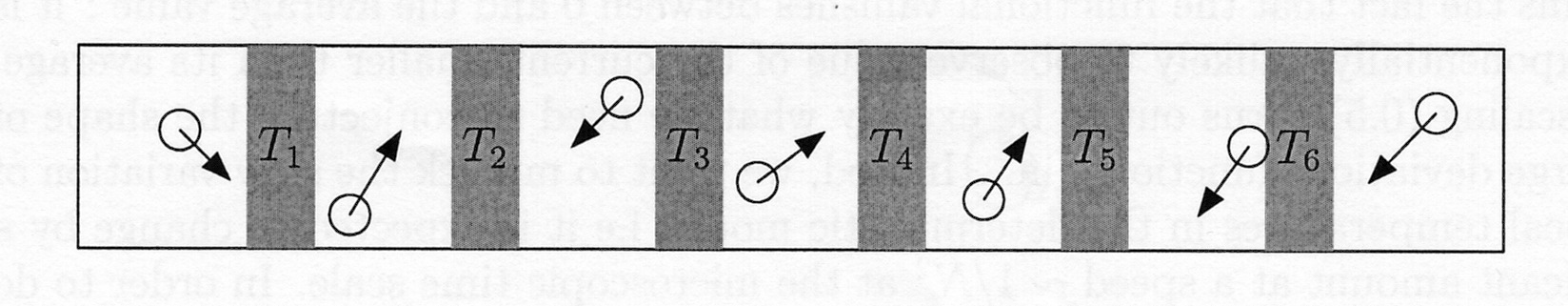} 
\label{thermalwalls}
\caption{The motion of a particle between two thermal walls}
\end{figure}

We delete the vertical dimension that is, for our purpose, irrelevant. And we first concentrate on the motion  of one particle between two thermal walls located at the ends of the interval $[0,1]$.  We denote its position and velocity by $(q(s), p(s))$.

\vspace{5mm}
\begin{figure}[thb]
\includegraphics[width = .50\textwidth]{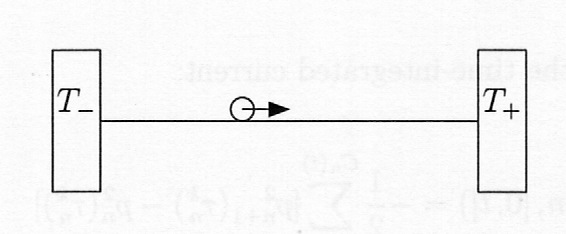} 
\label{2walls}
\caption{The motion of a particle between two thermal walls}
\end{figure}

\vspace{5mm}
The motion is ballistic inside the interval and when the particle encounters one wall, its velocity is reversed and its speed is updated with the law :
$$
\varphi_\pm(v)=\beta v e^{-\beta_{\pm} \frac{v^2}{2}},\quad \beta_\pm=T_\pm.
$$
One may associate to the motion of the particle the time-integrated current :
$$
J[0,t]=\sum_{k=1}^{N^-_t} \frac{(v^-_k)^2}{2}-\sum_{k=1}^{N^+_t} \frac{(v^+_k)^2}{2}
$$
where $v^\pm_k\sim \varphi^\pm(v)=\beta_\pm v e^{-\beta_\pm \frac{v^2}{2}}$ and $N^\pm_t$ counts the number of visits to the $\pm$-side of the interval.
Regarding this current, one can show the following :
$$
\lim_{t\to\infty}\frac{J[0,t]}{t}=\frac{T_--T_+}{(\frac{\pi}{2T_-})^{\frac 1 2}+(\frac{\pi}{2T_+})^{\frac 1 2}}\quad {\rm a.s.}
$$
where $T_-$ and $T_+$ are the left and right temperatures.
But one can go one step further and study the fluctuations of the LHS before the limit is taken.
Namely, one may compute the large deviations functional of the current $\cI(j,\tau,T)$ roughly defined as,
$$
\bbP_{\tau,T}\left (\frac{J[0,t]}{t}= j\right)\sim e^{-t \cI(j,\tau,T)},\quad t\to\infty.
$$
$\bbP_{\tau,T}$  is the stochastic dynamics with a fixed temperature difference $\tau=T_--T_+$ and average temperature $T=\frac{T_-+T_+}{2}$. 

If $\tau\ne0$ then one gets the following scaling result, 
\begin{equation}
\lim_{\varepsilon\downarrow 0}\varepsilon^{-2}\cI(\varepsilon j,\varepsilon\tau,T)={\mathcal G}(j,\tau,T)=\left\{\begin{array}{ll}
\frac{(j-\kappa\tau)^2}{4\kappa T^2} \ \ {\rm if} \ \ j\tau>\kappa\tau^2  \\
0  \ \ {\rm if} \ \ j\tau\in[0,\kappa\tau^2] 
\\ 
-\frac{j\tau}{2 T^2} \ \ {\rm if} \ \ j\tau\in[-\kappa\tau^2,0] 
\\
\\
\frac{j^2+\kappa^2\tau^2}{4\kappa T^2} \ \ {\rm if} \ \ j\tau<-\kappa\tau^2,
\end{array}
\right.
\label{scaling}
\nonumber
\end{equation}

\noindent  where $\kappa=(\frac{T}{2\pi})^{\frac 1 2}$.  
\begin{figure}[thb]
\includegraphics[width =.5\textwidth]{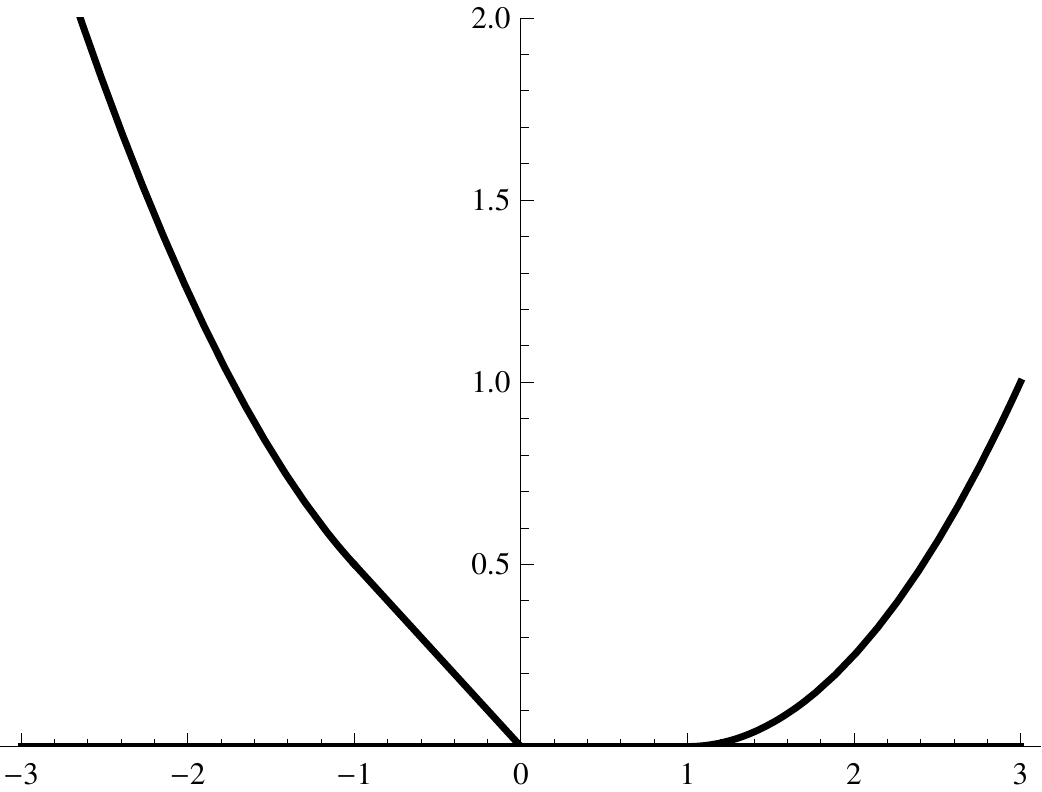} 
\caption{ Plot of ${\mathcal G}$ as a function of $j$ for $\kappa\tau=\kappa T^2=1$}
\end{figure}
The reason for the affine part of the functional is the fact that the particle will get a small velocity with a not so small probability and this will make the occurence of small values of $N_t$ (and thus of $J([0,t])$) quite likely.  This explains the fact that the functional vanishes between $0$ and the average value $\kappa\tau$ : it is {\it not} exponentially unlikely to observe values of the current smaller than its average.
The scaling expressed in the relation above turns out to be exactly what we need to conjecture the shape of the large deviations functional in the original billiard dynamics.
Indeed, we want to mimick the slow variation of the local temperatures in the deterministic model, i.e it is expected to change by a significant amount at a speed $\sim 1/N^2$ at the microscopic time scale.  In order to do so, we let the temperatures $\{T(n,t)\}_{0\leq n\leq N}$ of the scatterers in figure $3$ evolve such that they satisfy the equation
$$
T(n,t+N^2\Delta t)-T(n,t)=J(n,[t,t+N^2\Delta t])-J(n-1,[t,t+N^2\Delta t]),
$$
where the currents entering the equation are the ones on the LHS and RHS of the scatterer $n$.
Define $T_N:[0,1]\times \bbR^+\to \bbR^+$ and $\cJ_N:[0,1]\times \bbR^+\to \bbR$.
\begin{eqnarray}
T_N(x,t)&=&T([Nx],N^2t)\nonumber\\
\cJ_N(x,t)&=&N\cdot \frac{1}{N^2\Delta t}\cdot  J(n,[N^2t,N^2(t+\Delta t)]) ,
\nonumber
\end{eqnarray}

Then, one can show \footnote{Precise statements as well as rigorous proofs of the results below will be the subject of an upcoming publication} :
when $N\to\infty$, $\Delta t\to 0$, $(T_N,\cJ_N)$ converge in $L^2$  to the unique solution $(\hat T,\hat \cJ)$  of
\begin{equation}
\left\{\begin{array}{ll}
\partial_t \hat T(x,t)= -\partial_x \hat\cJ(x,t)\\
\hat \cJ(x,t)=-\kappa(\hat T(x,t))\partial_x \hat T(x,t)
\end{array}
\right.
\nonumber
\end{equation}
with $\kappa(T)=(\frac{T}{2\pi})^{\frac 1 2}$ and suitable b.c.

At finite $N$, and for each $(x,t)\in [0,1]\times [0,1]$, $\cJ_N(x,t)$ and $T_N(x,t)$ are random variables  and the object one is really interested in is the large deviations functional $\hat\cI$ appearing in :
\begin{equation}
\bbP\left(\{T_N\simeq \hat T,\cJ_N\simeq j\}\;{\rm on [0,1]\times[0,1]}\right)\sim \exp[-N \,\hat\cI(j,\hat T)].
\nonumber
\end{equation}
and one can show that
\begin{equation}
\hat \cI(j,\hat T)=\int_0^1 dt\int_0^1 dx\; \cG(j(x,t),\partial_x \hat T(x,t),\hat T(x,t))
\nonumber
\end{equation}
if $j$ and $\hat T$ satisfy $\partial_s\hat T(x,s)=-\partial_x j(x,s)$, and $\hat{\mathcal I}=+\infty$ otherwise.  The integrand $\cG$ is given by, 
\[
\cG(j,\tau,T)=\left\{\begin{array}{ll}
\frac{(j-\kappa\tau)^2}{4\kappa T^2} \ \ {\rm if} \ \ j\tau>\kappa\tau^2  \\
0  \ \ {\rm if} \ \ j\tau\in[0,\kappa\tau^2] 
\\ 
-\frac{j\tau}{2 T^2} \ \ {\rm if} \ \ j\tau\in[-\kappa\tau^2,0] 
\\
\\
\frac{j^2+\kappa^2\tau^2}{4\kappa T^2} \ \ {\rm if} \ \ j\tau<-\kappa\tau^2.
\end{array}
\right.
\]
And therefore,
\[
\cG(j,\tau,T)\neq\frac{(j-\kappa\tau)^2}{4\kappa T^2},
\]
which shows that local billiard dynamics should have macroscopic fluctuations properties very different from the ones observed in the usual stochastic interacting particles such as the exclusion process.  The large deviation functional is not a strictly convex function of the current.  This aspect should translate into some interesting phase transitions-like effects.  

The stochastic dynamics that is introduced is a very crude approximation of the deterministic dynamics.  However, one should note that it is not the fact that we replaced the deterministic collisions by stochastic ones that is responsible for the special behaviour of $\cG$. This is caused by what is left of the deterministic dynamics, namely the ballistic motion of the particle and the fact that one assumes the particles to be in local equilibrium. Therefore, one should expect this feature to be robust and independent of the approximations that were made for the collisional part of the dynamics.

\vspace{5mm}
\noindent {\bf Acknowledgments.} The author acknowledge the support of the French Ministry of education through the ANR grant SHEPI (2010-2013)

\end{document}